%% file: main.tex
\documentclass[a4paper,11pt]{article}

\usepackage{booktabs}
\usepackage{fourier}
\usepackage{color}
 \usepackage{graphicx}
\usepackage{url}
\usepackage[affil-it]{authblk}
\usepackage{amsmath}
\usepackage{wrapfig}

\usepackage[T1]{fontenc}
\usepackage{times}

\usepackage{array}
\usepackage{mdwmath}
\usepackage{mdwtab}
\usepackage{eqparbox}
\usepackage{multirow}
\usepackage{stmaryrd}
\usepackage{paralist}
\usepackage{dsfont}
\usepackage{verbatim}
\usepackage{calc}
\usepackage{amssymb}
\usepackage{amstext}
\usepackage{amsthm}
\usepackage{multicol}
\usepackage[small]{caption}
\usepackage{xcolor}
\usepackage{rotating}
\usepackage{subcaption}
\usepackage{adjustbox}

\usepackage{algorithm}
\usepackage{algorithmic}
\usepackage[numbers,sort&compress]{natbib}

\usepackage{hyperref}

\usepackage{xcolor}
\usepackage{pgfplots,pgfplotstable}
\usepackage{tikz}
\usepackage{lipsum,adjustbox}

\usepackage[autostyle]{csquotes}

\usetikzlibrary{calc,positioning}
\pgfplotsset{compat=newest}

\makeatletter
\newcounter{subhyp} 
\let\savedc@hyp\c@hyp

\newcommand{\normhyp}{%
  \let\c@hyp\savedc@hyp 
  \renewcommand\thehyp{\arabic{hyp}}%
}

\begin{document}
\title{Sub-Pixel Back-Projection Network For Lightweight Single Image Super-Resolution}


\author{Supratik Banerjee$^{1,2}$, Cagri Ozcinar$^{1}$, Aakanksha Rana$^{1}$, Aljosa Smolic$^{1}$, and  Michael Manzke$^{1}$}

\affil{$^{1}$School of Computer Science and Statistics, Trinity College Dublin, Ireland.\\$^{2}$Rawky Tech LLP, Mumbai, India.}

\thispagestyle{empty}
\date{}



%
%

%
%
\maketitle
%
\begin{abstract}
Convolutional neural network (CNN)-based methods have achieved great success for single-image super-resolution (SISR). However, most models attempt to improve reconstruction accuracy while increasing the requirement of number of model parameters. To tackle this problem, in this paper, we study reducing the number of parameters and computational cost of CNN-based SISR methods while maintaining the accuracy of super-resolution reconstruction performance. To this end, we introduce a novel network architecture for SISR, which strikes a good trade-off between reconstruction quality and low computational complexity. Specifically, we propose an iterative back-projection architecture using sub-pixel convolution instead of deconvolution layers. We evaluate the performance of computational and reconstruction accuracy for our proposed model with extensive quantitative and qualitative evaluations. Experimental results reveal that our proposed method uses fewer parameters and reduces the computational cost while maintaining reconstruction accuracy against state-of-the-art SISR methods over well-known four SR benchmark datasets.
\footnote{This publication has emanated from research conducted with the financial support of Science Foundation Ireland (SFI) under the Grant Number 15/RP/2776.} Code is available at \url{https://github.com/supratikbanerjee/SubPixel-BackProjection_SuperResolution}.


\end{abstract}

%
\input{introduction.tex}
\input{relatedwork.tex}
\input{proposedmethod.tex}

\input{results.tex}

\input{conclusion.tex}


\footnotesize
\bibliographystyle{IEEEbib}
\bibliography{refs}

\end{document}

%% file: introduction.tex
\section{Introduction}

Single image super-resolution (SISR) is the process of recovering the high-resolution (HR) image from a given low-resolution (LR) image~\cite{super_res_def_2003}. With the success in signal processing and machine learning, many learning-based SISR methods have been proposed in the literature, demonstrating promising results. Nowadays, these methods can be used in different applications~\cite{yang2014single, Wang2019-qi} such as medical imaging, surveillance, face recognition, and virtual reality~\cite{OzcinarRana2019}.

Given the advances in SISR, it remains a challenge to deploy the most existing SISR models in real-time applications, demanding compact deep neural network architectures. In particular, some emerging applications require faster SISR methods to boost the imaging performance. For example, modern graphic cards can raise a game's frame rates using SISR algorithm ~\cite{nvidiadlss}. In fact, most of the recent SISR algorithms are based upon very deep neural networks, requiring high number of parameters and computational cost for graphically-intensive workloads~\cite{yang_survey}. 

In this paper, we propose a new convolutional neural networks (CNNs)-based SISR method with an objective of factoring minimal reduction in perceptual quality while maintaining computational complexity. We use the previously developed SISR method in~\cite{DBPN}, and reduce its network parameters by simplifying the back-projection network architecture. For this, we replace the densely connected up- and down-projection units which comprise of several deconvolution and convolution layers by our proposed sub-pixel back-projection (SPBP) block. 
Experimental results validate the effectiveness of our proposed method in reconstructing accurate SR images. The proposed model requires a small number of parameters and low computational cost against several state-of-the-art SISR methods over four well-known SR test datasets. In addition, we demonstrate two smaller variations of our network, SPBP-S (small) and SPBP-M (medium), which use even fewer parameters and has significantly lower computational cost. 


The rest of the paper is organized as follows: Section~\ref{related-work} discusses the related CNN-based SISR works. Section~\ref{model} explains our proposed SISR model. Experimental results are presented in Section~\ref{results}. Finally, Section~\ref{conclusion} concludes the paper.

%% file: relatedwork.tex
\section{Related Work}
\label{related-work}




Inspired by the performance improvements obtained by CNNs on computer vision tasks such as image-to-image translation~\cite{rana}, image captioning~\cite{Ghosal_2019_ICCV}, Dong~\textit{et al.} proposed an SRCNN method~\cite{SRCNN}. This work proposed a three-layer network to learn the mapping between the desired HR image and its bicubic up-sampled LR image. 
Motivated by SRCNN, many CNN-based research works have been shown to use deeper networks to increase representation power further. For instance, Kim~\textit{et al.}~\cite{VDSR} presented a very deep SR (VDSR) architecture to significantly improve the SR image reconstruction accuracy with the use of a 20 layer VGG network \cite{VGG} along with global residual learning. 
Recently, Haris \textit{et al.}~\cite{DBPN}~proposed a deep back-projection network (DBPN), which was based on the idea of iterative up- and down- sampling. 
However, their proposed network uses large filter sizes which increases the number of parameters, leading to higher computational complexity. Ahn~\textit{et al.}~\cite{CARN} designed a cascading mechanism on residual networks, which effectively boost the performance with multi-level representation and multiple short-cut connections for learning residuals in LR feature space.  Li~\textit{et al.}~\cite{SRFBN} proposed (SRFBN) to improve reconstruction performance while having low parameters to reduce chances of over-fitting using a feedback mechanism, but it increases the computational cost of the network.


Computational efficiency of the neural networks designed for SISR is important. Dong \textit{et al.}~\cite{FSRCNN}, for instance, designed an efficient network structure for fast SISR, called fast SR CNN (FSRCNN). 
With a similar aim, Shi \textit{et al.}~\cite{ESPCN} proposed an efficient sub-pixel CNN (ESPCN). In their work, pixel shuffle network was used to upscale the image at the final step of the SR process. Even though their network demonstrates real-time performance, it lacks high reconstruction quality due to its architectural simplicity. Recently, a few SR networks \cite{CARN,IDN,FLSR} have been proposed to have low parameters and low computational complexity, while maintaining state-of-the-art reconstruction performance.  

%% file: proposedmethod.tex
\section{Proposed Model}
\label{model}

\begin{figure}[t]
    \centering
    \includegraphics[width=0.7\linewidth]{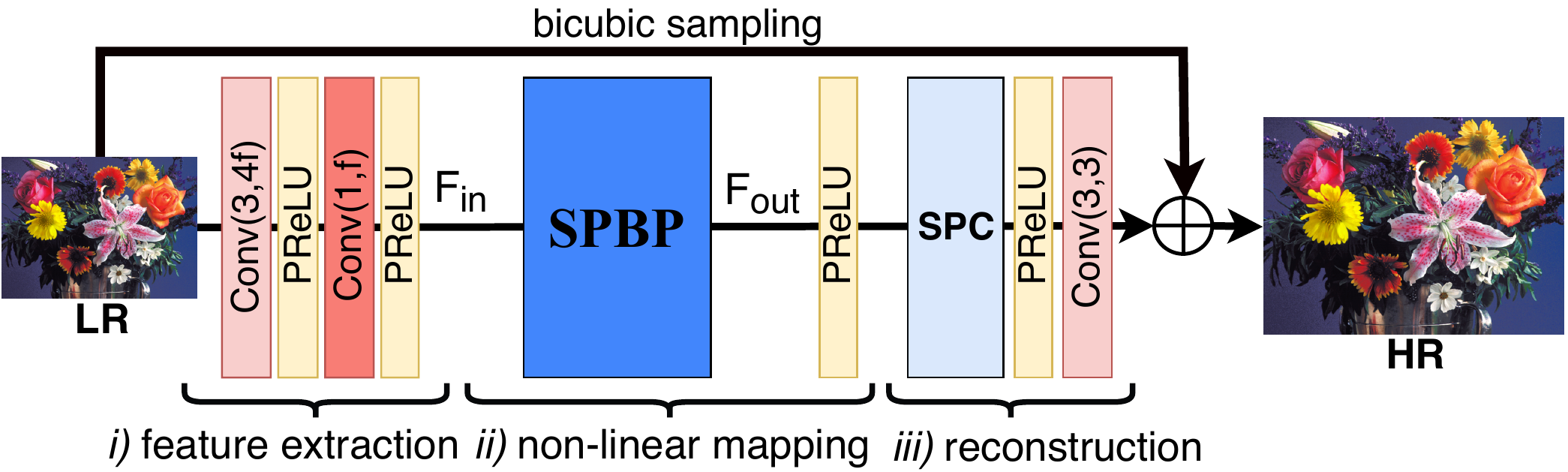}
    \caption{Proposed network architecture for SISR.}

    \label{fig:network}
\end{figure}

As shown in Fig.~\ref{fig:network}, our proposed network  architecture consists of three main blocks, namely, \textit{i}) feature extraction (FE), \textit{ii}) non-linear mapping (NLM), and \textit{iii}) reconstruction. At the first block, we extract shallow features from the LR image. The second block extracts deeper features using an iterative back-projection technique. The third block up-samples and refines the final SR image.
In the following, we present details of each block where convolutions are denoted as $Conv(k, n)$ with $k$ being the filter size and $n$ being the number of filters. 

\subsection{Feature extraction}
The FE block consists of two convolution layers with PReLU as activation layers, similar to the architectures proposed in~\cite{DBPN, FSRCNN}. The FE block is defined as:
 \begin{equation}
     {F}_{in}^{0} = {C}_{0}^{FE}(I_{LR}), \quad
     \text{and} \quad
     {F}_{in}^{1} = {C}_{1}^{FE}({F}_{in}^{0}),
 \end{equation}
where ${C}_{0}^{FE} = Conv(3, 4f)$ with ${C}_{1}^{FE} = Conv(1, f)$, and $f$ is the base number of filters. The low-level representation, ${F}_{in}^{0}$, is obtained from the LR image, $I_{LR}$, and the refined feature ${F}_{in}^{1}$ is obtained by ${F}_{in}^{0}$.

\subsection{Non-linear mapping}
Next, we present details about our proposed NLM block, called SPBP. Here, we reduce the computational cost for SISR, building upon and simplifying the back-projection block developed in~\cite{DBPN}. This method, called DBPN, proposes the use of densely connected up and down projection units. These units make use of multiple convolution and deconvolution (Dconv) layers to back-project the feature maps, which makes the network computationally expensive.

To reduce the model complexity of DBPN, we propose to replace these up- and down-projection units and their error feedback mechanism with up- and down-sampling layers as SPC and convolution layers. 
Our inspiration for this new approach of using SPC over Dconv is based on the work of Shi \textit{et al.}~\cite{ESPCN}, where it is described that the SPC layer is $\log_{2}r^2$ times faster than Dconv layer in the forward pass. Since SPC operates in LR space on a feature map of size $\left(n, \frac{W}{s}, \frac{H}{s}\right)$ and Dconv layer operates in HR space on a feature map of size $\left(\frac{n}{s^2}, W, H\right)$, where $W$ and $H$ are the dimensions of the input. We can represent the information contained in its feature maps as: $SPC = {LR}\left(n \times \frac{W}{s} \times \frac{H}{s}\right)$ and $Dconv = {HR}\left(\frac{n}{s^2} \times W \times H\right)$.
The complexity of the layers with a filter size of $k \times k$ and scaling factor $s$ will then be:
\vspace{-0.5em}
\begin{equation}
SPC = {O}\left(n \times n \times k \times k \times \frac{W}{s} \times \frac{H}{s}\right)
\end{equation}
\begin{equation}
Dconv = {O}\left(\frac{n}{s^2} \times \frac{n}{s^2} \times sk \times sk \times W \times H\right)
\end{equation}
Thus, the number of parameters are:
\begin{equation}
\label{eq6}
SPC = {LR}\left(n \times n \times k \times k\right)
\end{equation}
\begin{equation}
\label{eq7}
Dconv = {HR}\left(\frac{n}{s^2} \times \frac{n}{s^2} \times sk \times sk\right)
\end{equation}
For the same information retention and computational complexity, as shown in Eqs. \eqref{eq6} and \eqref{eq7}. 
SPC contains larger number of parameters compared to Dconv, and therefore, upholds a higher representation power without adding computational complexity. For this reason, we propose to use SPC in order to reduces network parameters by simplifying the back-projection network architecture. This approach provides higher representation power and achieves an efficient feature mapping.

\begin{figure}[ht]
     \centering
     \includegraphics[width=0.65\linewidth]{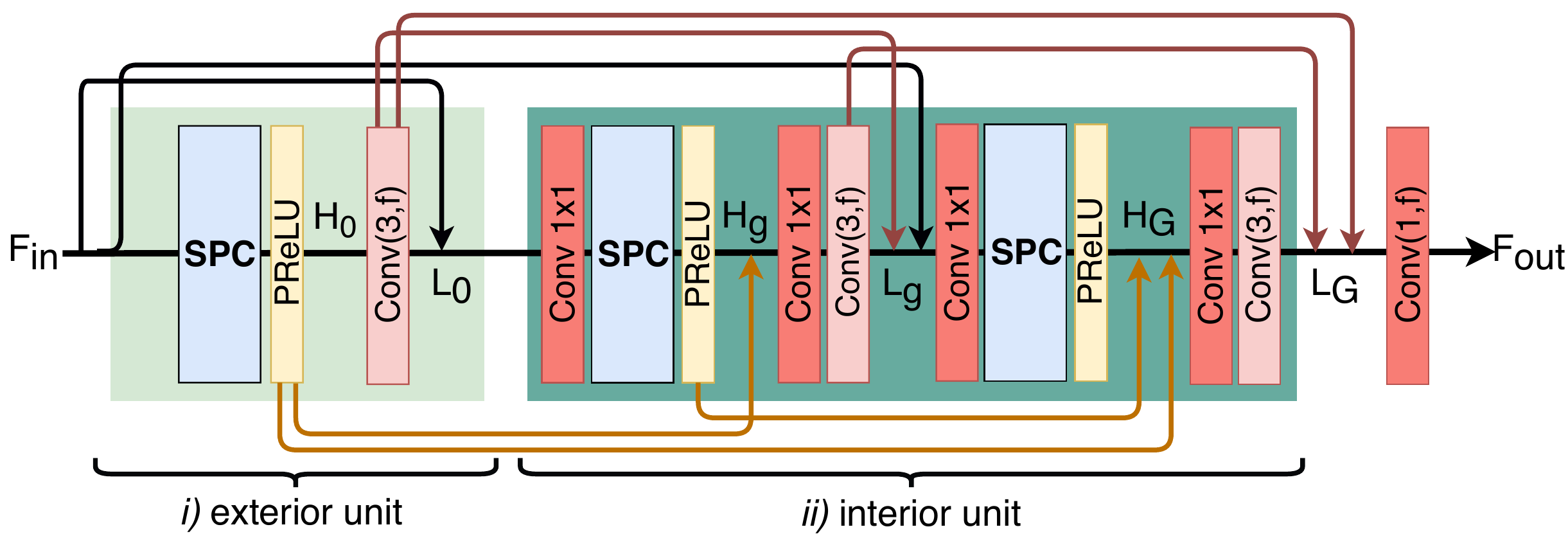}
     \caption{ Sub-Pixel Back-Projection Block }
\label{fig:SPBP}
\end{figure}

Figure~\ref{fig:SPBP} shows the design of SPBP, which comprises of an exterior and interior unit.
The exterior unit is defined as:
  \begin{equation}
     {H}_{0} = {PS}({C}_{0,0}^{NLM}({F}_{in}^{1})\uparrow_{s}, 
     \text{ and } 
     {L}_{0} = {C}_{0,1}^{NLM}({H}_{0}) \downarrow_{s},
 \end{equation}
 where $\uparrow_{s}$, $\downarrow_{s}$ represent up-sample and down-sample operations respectively with a scale factor $s$. Also, ${C}_{0,0}^{NLM}$ represents $Conv(3, fs^2)$, where ${PS}$ is the pixel-shuffle layer, which defines SPC. The SPBP block takes ${F}_{in}^{1}$, which is the first LR feature map in this block as input and produces an HR feature map, ${H}_{0}$. This is back-projected to a LR feature map ${L}_{0}$ using ${C}_{0,1}^{NLM}$, which represents $Conv(3, f)$. This is a single group of the proposed SPBP block.

 

The use of DenseNet~\cite{densenet} has demonstrated the alleviation of vanishing gradient problem. Also, the use of dense skip connections help to generate powerful high-level representations and encourages feature reuse. Inspired by this, we introduce the use of dense connections in SPBP block, as similar to \cite{DBPN}, which forms the interior unit of the block. 
Thus the interior unit of $G$ groups is formulated as:
\begin{equation}
    {H}_{g}={PS}({C}_{g,0}^{NLM}\left(\left[{F}_{in}^{1}, {L}_{0}, \ldots, {L}_{g-1}\right])\right)\uparrow_{s},
\end{equation}
\begin{equation}
    {L}_{g}={C}_{g,1}^{NLM}\left(\left[{H}_{0}, {H}_{1}, \ldots, {H}_{g}\right]\right) \downarrow_{s},
\end{equation}
\begin{equation}
    {F}_{out}={C}_{out}\left(\left[{L}_{1}, {L}_{2}, \ldots, {L}_{G}\right]\right).
\end{equation}
where $[{F}_{in}^{1}, {L}_{0}, \ldots, {L}_{g-1}]$ refers to the concatenation of ${F}_{in}^{1}$, LR feature maps $0,...,g-1$ and  ${H}_{g}$ and is the HR feature map produced by the up-projection layer in the $g^{th}$ group. Similarly, $[{H}_{0}, {H}_{1}, \ldots, {H}_{g}]$ refers to the concatenation of HR feature maps  $0,...,g$ and  ${L}_{g}$ is the LR feature map produced by the down-projection layer in the $g^{th}$ group. ${C}_{out}$ is a compression unit representing $Conv(1, f)$ to generate the output ${F}_{out}$ by fusing LR features from the previous levels $1,...,G$ of the SPBP block.


\subsection{Reconstruction}
This block uses a SPC layer which up-scales the LR feature map obtained from the SPBP block. This is followed a convolution layer which refines the up-sampled feature map. The reconstruction layer is defined as:
  \begin{equation}
     {I}_{0}^{Res} = {PS}({C}_{0}^{R}({F}_{out})\uparrow_{s},
 \text{ and } 
     {I}_{1}^{Res} = {C}_{1}^{R}({I}_{0}^{Res}),
 \end{equation}
   \begin{equation}
     {I}_{SR} = {I}_{1}^{Res} + f_{UP}({I}_{LR}),
 \end{equation}
where ${I}_{0}^{Res}$ is the residual upscale of ${PS}({C}_{0}^{R}({F}_{out}))$ with input ${F}_{out}$. ${I}_{1}^{Res}$ is the refined residual HR feature map derived from ${C}_{1}^{R}(.)$, which is  a $Conv(3 ,f_{out})$ where, $f_{out} = 3$ is the output feature map ``RGB". Inspired by \cite{nvidiagwmt, SRFBN} the super-resolved image is constructed by adding the refined HR feature map with $f_{UP}(.)$, which is bicubic up-sample of the LR image. Since the LR image contains abundant low-frequency information~\cite{RCAN}, this allows the network to bypass the LR information and focus only on the residual component from the HR image.


 

 

%% file: results.tex
\section{Results}
\label{results}

In this section, we first describe our training details, and then we evaluate our proposed SISR method with state-of-the-art SISR methods using quantitative and qualitative experiments.

\subsection{Training Details}

All experimentation was carried out on $\times2$ scaling factor between LR and HR. The LR images were obtained by down-sampling HR images from the training set of \textit{DIV2K}~\cite{DIV2K} dataset with bicubic interpolation. For training, the LR image-crop size was set as $48\times48$ with $40$ random crops per image. The mini-batch size was set to $40$ for all network configurations. Each proposed model was trained using the ADAM optimizer with L1 loss for $1000$ epochs, with $\beta_{1} = 0.9$, $\beta_{2} = 0.999$. The learning rate was initialized as $10^{-4}$ and decayed by a factor of $2$ in every $200$ epochs. Image augmentation was was used for training by randomly flipping horizontally or vertically and rotating the training images like \cite{EDSR, SRFBN}. Three different settings for the proposed SISR model, SPBP-S (small), SPBP-M (medium) and SPBP-L (large) use $(f=16, G=1)$, $(f=16, G=10)$, and $(f=32, G=10)$ configurations, respectively. The proposed models have been implemented using the PyTorch library~\cite{paszke2017automatic}. The training was performed using NVIDIA Titan-Xp GPU with 12 GB memory on Intel core i7-7700 machine.

\subsection{Evaluation}

To validate our proposed SPBP method, we performed a thorough experimental analysis using nine CNN-based state-of-the-art SISR algorithms: SRCNN~\cite{SRCNN}, FSRCNN~\cite{FSRCNN}, ESPCN~\cite{ESPCN}, VDSR~\cite{VDSR}, DBPN-SS~\cite{DBPN}, CARN~\cite{CARN}, IDN~\cite{IDN}, SRFBNs~\cite{SRFBN}, FLSR~\cite{FLSR}. As our focus is to develop a lightweight network for SISR, for simplicity, we do not show results for the published networks which are known to have a more complex model than CARN~\cite{CARN}. Each model was tested with four datasets, namely, \textit{Set5}~\cite{Yang_image}, \textit{Set14}~\cite{set14}, \textit{BSDS100}~\cite{arbelaez2010contour}, and \textit{Urban100}~\cite{Urban}.

In the following, we compare the performance between our proposed methods (SPBP-S, SPBP-M, SPBP-L, SPBP-L+), and state-of-the-art SISR methods using quantitative and qualitative analysis. Similar to other SISR methods \cite{EDSR, SRFBN, FLSR}, we applied the self-ensemble strategy during testing on SPBP-L to further improve the reconstruction performance, we denote this method as SPBP-L+.

 \begin{table}[!ht]
 	\caption{Quantitative Results on four datasets. The highest reconstruction accuracy is indicated in \textbf{\textcolor{red}{red}} and second highest reconstruction accuracy in \textit{\textcolor{blue}{blue}}. [$\times2$ upscaling]
	}
	\begin{adjustbox}{width=\columnwidth,center}
		\begin{tabular}{l c c c|c||c|c||c|c||c|c}
\toprule
& &
&\multicolumn{8}{c}{\textbf{Datasets}}\\

			\cmidrule(l){4-11}
\multirow{2}{*}{\textbf{Methods}} & \multirow{2}{*}{\textbf{\# of parameters}} &\multirow{2}{*}{\textbf{Multi-Adds}} &
			\multicolumn{2}{c}{\textit{Set5}}
			&\multicolumn{2}{c}{\textit{Set14}}
			&\multicolumn{2}{c}{\textit{BSDS100}}
			&\multicolumn{2}{c}{\textit{Urban100}}
			\\
			\cmidrule(l){4-5}
			\cmidrule(l){6-7}
			\cmidrule(l){8-9}
			\cmidrule(l){10-11}
&
			&
			& \multicolumn{1}{c|}{\underline{PSNR}} & \multicolumn{1}{c||}{\underline{SSIM}} & \multicolumn{1}{c|}{\underline{PSNR}} & \multicolumn{1}{c||}{\underline{SSIM}}  &
			\multicolumn{1}{c|}{\underline{PSNR}} & \multicolumn{1}{c||}{\underline{SSIM}}  &\multicolumn{1}{c|}{\underline{PSNR}} & \multicolumn{1}{c}{\underline{SSIM}} \\	
			SRCNN~\cite{SRCNN} & 69K & 63.8G & 36.66  & 0.9542 &  32.42 &  0.9063 &  31.36 &  0.8879 &  29.50 &  0.8946 \\
			FSRCNN~\cite{FSRCNN} & 25K & 15G &  37.00  & 0.9558 & 32.63 &  0.9088 &  31.53 &  0.8920 & 29.88 &  0.9020 \\	
			ESPCN~\cite{ESPCN} & 26K & 6.17G &  36.69  & 0.9547 & 32.50 &  0.9076 &  31.31 &  0.8882 & 29.35 &  0.8937 \\
			VDSR~\cite{VDSR} & 665K & 612.6G & 37.53  & 0.9587 & 33.03 &  0.9124 &  31.90 &  0.8960 & 30.76 &  0.9140 \\
			DBPN-SS~\cite{DBPN} & 109K & 66.2G & 37.44  & 0.9589 & 33.03 &  0.9127 &  31.81 &  0.8951 & 30.67 &  0.9128 \\
			CARN~\cite{CARN} & 960K & 223.7G &  37.76  & 0.9590 & 33.52 &  0.9166 &  32.09 &  0.8978 & \textit{\textcolor{blue}{31.92}} &  0.9256 \\	
			IDN~\cite{IDN} & 591K & 138.3G &  37.83  & 0.9600 & 33.30 &  0.9148 & 32.08 &  0.8985 &  31.27 &  0.9196 \\
			SRFBNs~\cite{SRFBN} & 282K & 679.7G &  37.82  & 0.9598 &  33.38 &  0.9155 &  32.08 &  0.8983 &  31.65 &  0.9232\\	
			FLSR~\cite{FLSR} & 717K & 271.4G & 37.79  & 0.9595  &  33.16 &  0.9143 & 32.06 &  0.8983 &  31.723 &  0.9183-\\

\midrule
			SPBP-S & 24K & 5.9G &  37.23  & 0.9577  &  32.85 &  0.9109 & 31.66 &  0.8930 & 30.37 &  0.9091 \\
			SPBP-M & 159K & 46.6G & 37.72  & 0.9593 &  33.33 &  0.9151 & 32.02 &  0.8975 & 31.43 &  0.9211 \\
			SPBP-L & 629K & 184G &  \textit{\textcolor{blue}{37.95}}  & \textit{\textcolor{blue}{0.9603}} &  \textit{\textcolor{blue}{33.54}} &  \textit{\textcolor{blue}{0.9171}} &\textit{\textcolor{blue}{32.15}} &  \textit{\textcolor{blue}{0.8994}} & 31.89 &  \textit{\textcolor{blue}{0.9262}} \\	
			SPBP-L+ & 629K & 184G &  \textbf{\textcolor{red}{38.05}}  & \textbf{\textcolor{red}{0.9606}} &  \textbf{\textcolor{red}{33.62}} &  \textbf{\textcolor{red}{0.9178}} &  \textbf{\textcolor{red}{32.21}} &  \textbf{\textcolor{red}{0.9001}} & \textbf{\textcolor{red}{32.07}} &  \textbf{\textcolor{red}{0.9277}}\\

\bottomrule
		\end{tabular}
\end{adjustbox}   
	\label{tab:quantresults}
\end{table}

\begin{figure*}[!ht]
\centering
\begin{minipage}[h]{.32\linewidth}
	\centerline{\includegraphics[width=\linewidth]{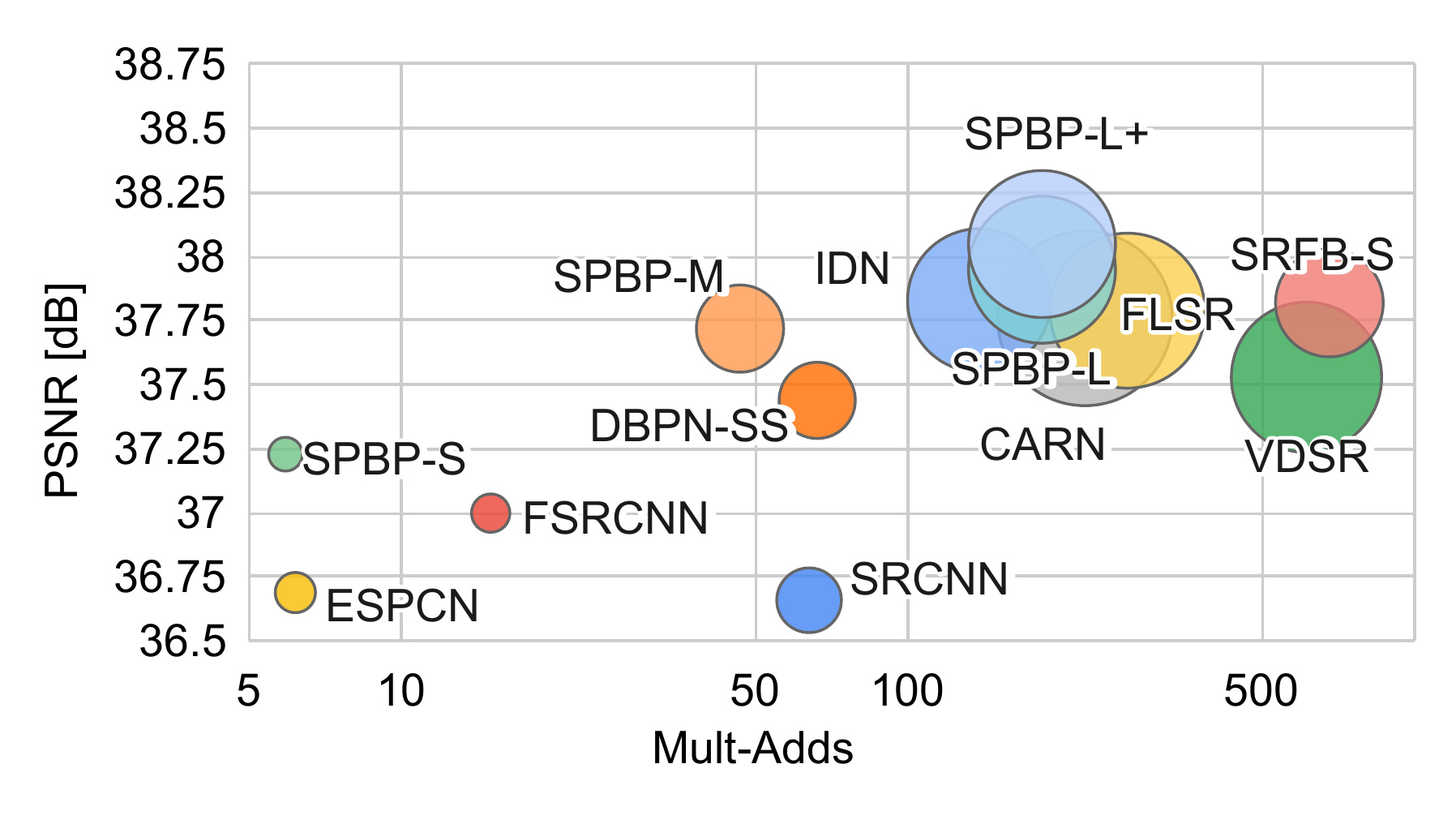}}
	\centerline{(a)~\textit{Set5}~\cite{Yang_image}}
\end{minipage}
\begin{minipage}[h]{.32\linewidth}
	\centerline{\includegraphics[width=\linewidth]{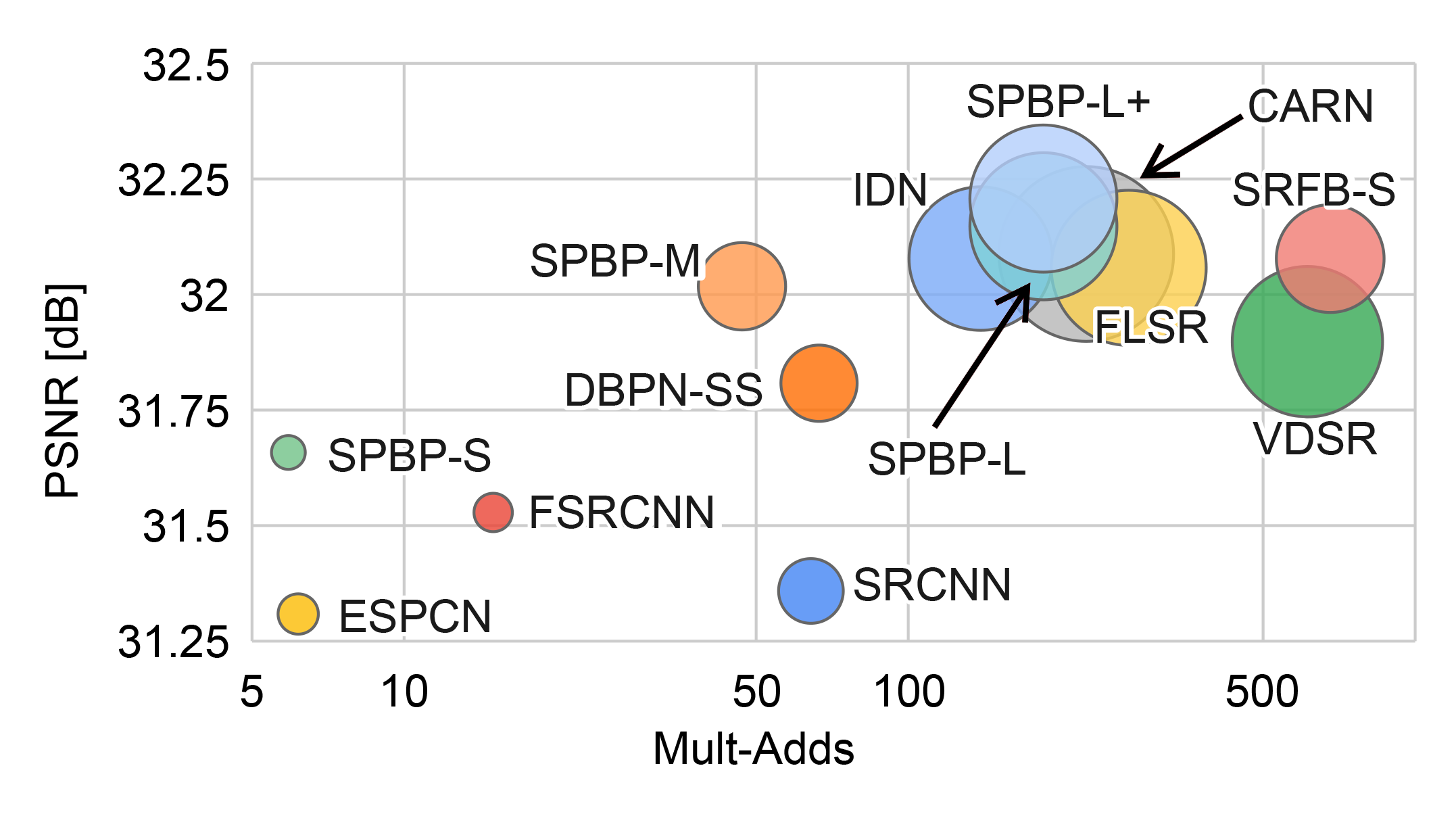}}
	\centerline{(b)~\textit{BSDS100}~\cite{arbelaez2010contour}}
\end{minipage}
\begin{minipage}[h]{.32\linewidth}
	\centerline{\includegraphics[width=\linewidth]{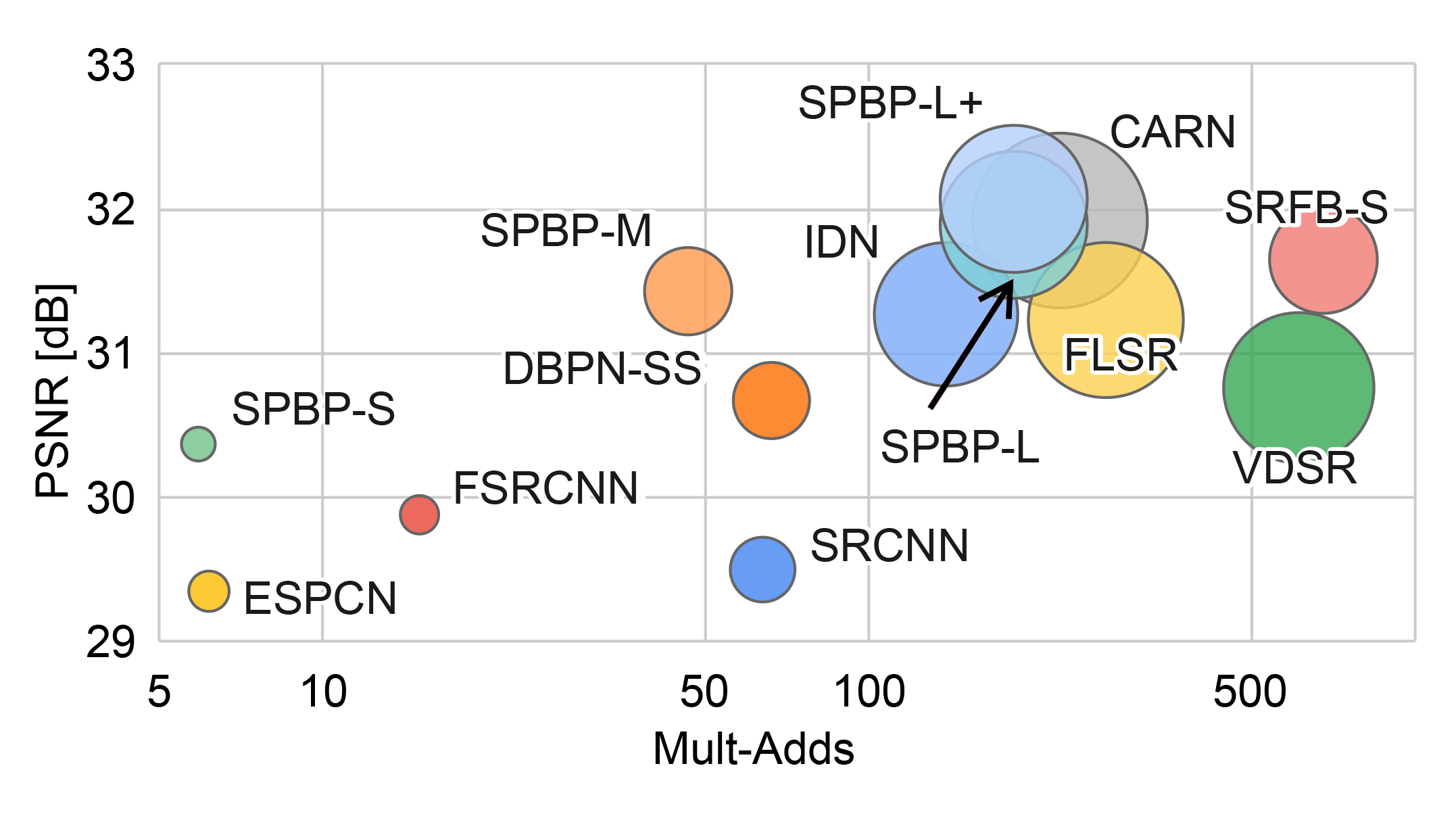}}
	\centerline{(c)~\textit{Urban100}~\cite{Urban}}
\end{minipage}
        \caption{Trade-off between reconstruction accuracy versus number of operations and parameters on three datasets. The $x$-axis and the $y$-axis denote the Multi-Adds and PSNR [dB],  and the size of the circle represents the number of parameters. The Mult-Adds is computed for HR image of size 720p.}
\label{performance}
\end{figure*}

\begin{figure}[!ht]
        \centering
        \includegraphics[scale=0.48]{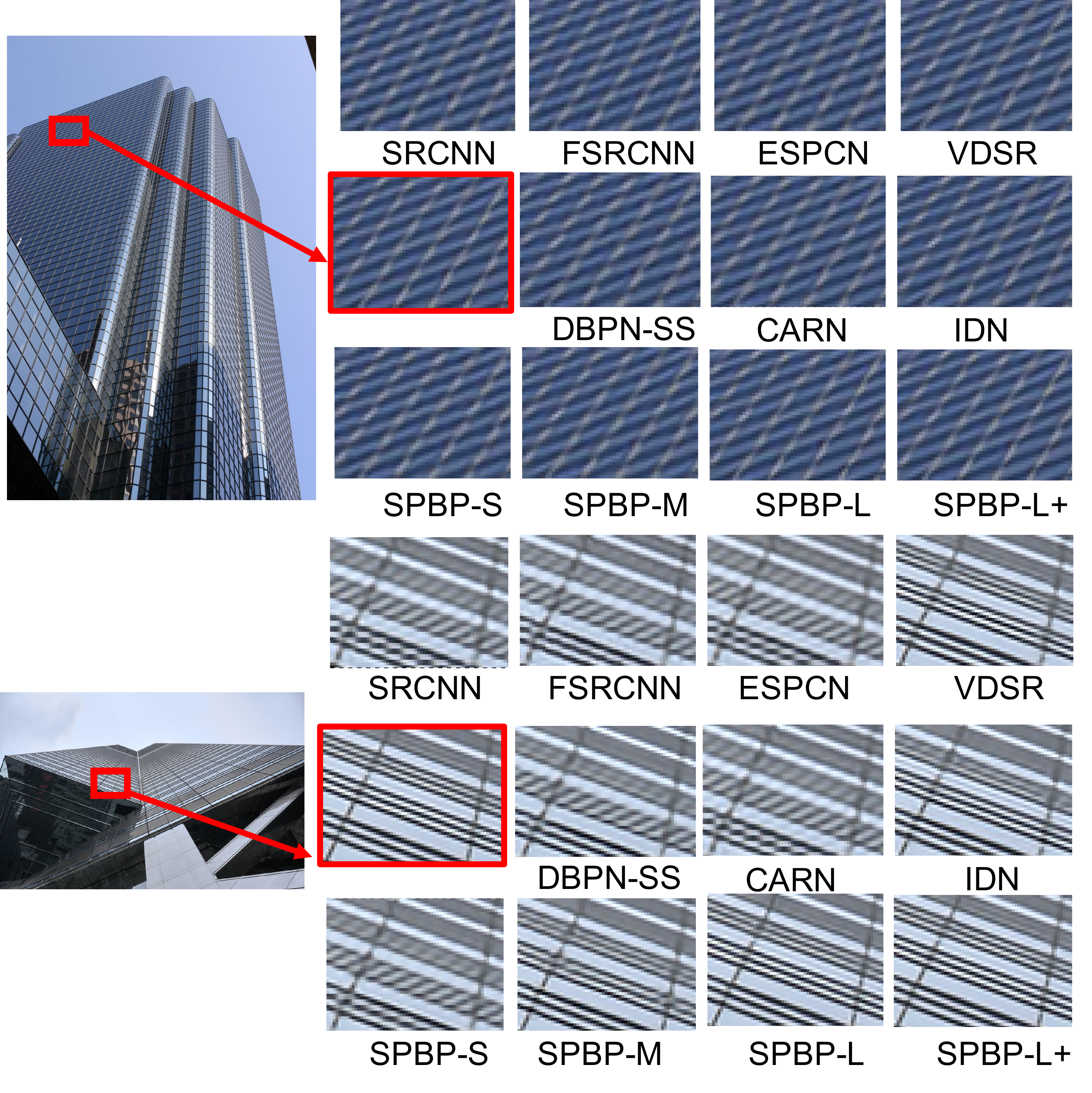}
        \caption{Qualitative comparison of our SPBP models with other works on ``img\_074" and ``img\_059" example images from the \textit{Urban100}.} 
 \label{performance_visual_1}   
\end{figure}

\subsubsection{Quantitative}

We measured the performance of each method for its reconstructed accuracy of the SR image using PSNR and SSIM. Here, similar to previous works~\cite{EDSR, VDSR}, we cropped $2$ pixels near image boundary and estimated quality scores using only the luminance channel (Y) of images. Also, we measure the computational complexity in terms of the number of operations with Mult-Adds, which is the number of composite multiply-accumulate operations. Table~\ref{tab:quantresults} compares the performance of the proposed SPBP-S, SPBP-M, and SPBP-L models with state-of-the-art methods in terms of $\#$ of parameters, computational complexity, and objective quality metrics.

We also examined the computational complexity of our model in comparison to other state-of-the-art methods concerning PSNR over the datasets. Fig.~\ref{performance} shows trade-off between reconstruction accuracy (in terms of PSNR) versus number of operations and parameters over three datasets: (a)-\textit{Set5}, \textit{BSDS100}, and \textit{Urban100}. In the experiment, the calculations were performed for HR image of size $720p$ ($1280\times720$). Looking at the results, we see that our proposed models (SPBP-S, SPBP-M, SPBP-L, and SPBP-L+) outperform state-of-the-art methods in terms of PSNR for comparable parameter size and has a much lower computational cost.

Overall, our SPBP-L+ model, which has nearly 629K parameters, shows the best reconstruction accuracy performance in most of the benchmark datasets in terms of objective quality scores. Further, we observe that SPBP-M which has only 159K parameters performs very close in most of the benchmark datasets to FLSR, SRFBN, IDN and CARN, all of which have about double or more parameters. Comparing models with less than 100K parameters, we can clearly see SPBP-S outperforms all existing models (SRCNN, FSRCNN, ESPCN). These results prove that our developed models handle the image feature better than the other state-of-the-art methods with fewer parameters and lower computational complexity.\\

\subsubsection{Qualitative}

To provide qualitative visual comparison between methods, Fig. \ref{performance_visual_1} shows some examples of reconstructed images from the \textit{Urban100} dataset. We see that the proposed model can construct HR images with higher quality, compared to most of the state-of-the-art methods. Also, we observe that the proposed models have visually similar or better results compared to other state-of-the-art networks, such as CARN, IDN, VDSR, but with lower parameters and computational expense. Especially, the proposed SPBP models construct high frequency patterns with subjectively closer to the original HR. 







%% file: conclusion.tex
\vspace{-1 mm}
\section{Conclusion}
\label{conclusion}
\vspace{-2 mm}
In this paper, we proposed a novel sub-pixel convolution-based dense iterative back-projection network architecture for single-image super-resolution tasks. We showed the reconstruction accuracy and computational efficiency of employing our proposed models (SPBP-S, SPBP-M,  SPBP-L) in terms of model parameters, quantitative quality measures (in terms of PSNR, SSIM), and qualitative evaluations. We also compared our proposed model with nine state-of-the-art SISR methods over well-known SR datasets and demonstrated that our proposed approach provides lower computational complexity while maintaining high reconstruction performance. This can be very well observed with SPBP-S which stands out to be the best performing network under 100K parameters.

